\documentclass[aps,prd,nofootinbib,twocolumn,superscriptaddress]{revtex4-1}

\usepackage[english]{babel} 
\usepackage{graphicx} 
\usepackage{commath}
\usepackage{chngcntr}
\usepackage[usenames,dvipsnames,svgnames,table]{xcolor}
\usepackage{bm}
\usepackage{soul}

\usepackage[colorlinks = true,
            linkcolor = blue,
            urlcolor = blue,
            citecolor = blue,
            anchorcolor = blue]{hyperref}

\def\omb{\Omega_{\mathrm{b}}h^2}
\def\omc{\Omega_{\mathrm{c}}h^2}
\def\sumnu{\Sigma m_\nu}
\def\fsig{f\sigma_8}
\def\sig{\sigma_8}

\def\am{\alpha_M}
\def\ab{\alpha_B}
\def\ak{\alpha_K}

\def\bea{\begin{eqnarray}}
\def\eea{\end{eqnarray}}

\begin{document}

\title{Improvements in cosmological constraints from breaking growth degeneracy}

\author{Louis Perenon}
\email{perenon.louis@yahoo.fr, corresponding author}
\affiliation{Department of Physics \& Astronomy, University of the Western Cape, Cape Town 7535, South Africa}

\author{St\'ephane Ili\'c}
\affiliation{Universit\'e PSL, Observatoire de Paris, Sorbonne Universit\'e, CNRS, LERMA, F-75014, Paris, France}
\affiliation{CEICO, Institute of Physics of the Czech Academy of Sciences, Na Slovance 2, Praha 8, Czech Republic}
\affiliation{IRAP, Universit\'e de Toulouse, CNRS, CNES, UPS, Toulouse, France}

\author{Roy Maartens}
\affiliation{Department of Physics \& Astronomy, University of the Western Cape, Cape Town 7535, South Africa}
\affiliation{Institute of Cosmology \& Gravitation, University of Portsmouth, Portsmouth PO1 3FX, UK}

\author{Alvaro de la Cruz-Dombriz}
\affiliation{Cosmology \& Gravity Group, Department of Mathematics \& Applied Mathematics, University of Cape Town, Rondebosch 7701, Cape Town, South Africa}

\begin{abstract}
The key probes of the growth of large-scale structure are its rate $f$ and amplitude $\sigma_8$. Redshift space distortions in the galaxy power spectrum allow us to measure only the combination $f\sigma_8$, which can be used to constrain the standard cosmological model or alternatives. By using measurements of the galaxy-galaxy lensing cross-correlation spectrum or of the galaxy bispectrum, it is possible to break the $f\sigma_8$ degeneracy and obtain separate estimates of $f$ and $\sigma_8$ from the same galaxy sample. Currently there are only a handful of such separate measurements, but even this allows for improved constraints on cosmological models. We explore how having a larger and more precise sample of such measurements in the future could constrain further cosmological models. We consider what can be achieved by a future nominal sample that delivers a $\sim 1\%$ constraint on $f$ and $\sigma_8$ separately, compared to the case with a similar precision on the combination $f\sigma_8$. For the six cosmological parameters of $\Lambda$CDM, we find improvements of $\sim\! 5$--$50\%$ on their constraints. For modified gravity models in the Horndeski class, the improvements on these standard parameters are $\sim\! 0$--$15\%$. However, the precision on the sum of neutrino masses improves by 65\% and there is a significant increase in the precision on the background and perturbation Horndeski parameters.
\end{abstract}

\maketitle

\section{Introduction}

The growth of large-scale structure is sensitive to the theory of gravity and its measurement is a powerful test of the standard and alternative models of cosmology. It is characterised at the most basic level by the rate of growth $f=-d\ln D/d\ln(1+z)$, where $D(z)$ is the growth function of the linear matter density contrast, $\delta(z,\bm{k})=D(z)\delta(z_{\rm in},\bm{k})/D(z_{\rm in})$, given an initial redshift $z_{\rm in}$. This rate governs the evolution of peculiar velocities, whose impact on the observed galaxy power spectrum is to introduce a redshift space distortion (RSD). Measurement of this anisotropy at redshift $z$ delivers an estimate of $f(z)\sigma_8(z)$, where $\sigma_8$ fixes the amplitude of the matter density fluctuations. The degeneracy between $f$ and $\sigma_8$ echoes the degeneracy between the linear galaxy bias and $\sigma_8$, and it cannot be broken via RSD power spectrum measurements alone.

The degeneracy can be broken by using an alternative observable in the galaxy sample that involves $\sig$ or $f$. For example, combining RSD power spectrum measurements with galaxy-galaxy lensing measurements has produced separate estimates of $f$ and $\sig$ \cite{delaTorre:2016rxm,Shi:2017qpr,Jullo:2019lgq}. There are currently only a handful of such estimates, but even with only three separated data pairs, constraints on cosmological models improve noticeably \cite{Perenon:2019dpc}. Another way to break the degeneracy is by combining RSD measurements in the power spectrum and bispectrum \cite{Gil-Marin:2016wya}.

Breaking the growth degeneracy is expected to break degeneracies between certain cosmological and modified gravity parameters. Here we confirm this expectation by computing the improvement in precision when using future separated measurements of $f$ and $\sig$ as compared to using the usual combined measurements $f\sig$. We make forecasts for the standard $\Lambda$CDM model and for scalar-tensor theories in the Horndeski class \cite{Horndeski:1974wa}, using the effective field theory (EFT) of dark energy \cite{Gubitosi:2012hu,Bloomfield:2012ff} (see \cite{Frusciante:2019xia} for a recent review and \cite{Gleyzes:2015rua,Alonso:2016suf,Leung:2016xli,Abazajian:2016yjj,Reischke:2018ooh,Mancini:2018qtb,Frusciante:2018jzw,Ballardini:2019tho} for more general Horndeski forecasts).

\section{Models}\label{sec:Models}

We consider two models to assess the constraining power of the different growth of structure quantities. The first is the standard cosmological model $\Lambda$CDM, whose free parameters are \cite{Aghanim:2019ame}
\begin{equation}
\big\lbrace \omb,\omc, H_0, \tau,A_{\rm s}, n_{\rm s},\sumnu \big\rbrace ,
\end{equation}
where the total neutrino mass $\sum m_\nu$ is equally shared by the three degenerate species.

For the second, we chose the popular benchmark for studies of alternative gravitational models \cite{Frusciante:2019xia} that are Horndeski theories \cite{Horndeski:1974wa}. They are the most general covariant scalar-tensor theories with direct second-order equations of motion. We use in particular their description of linear perturbations provided by the $\alpha$-EFT basis \cite{Bellini:2014fua}. See \cite{Bellini:2014fua} for complete details of the construction of the action.

Observations suggest that the speed of gravitational waves is equal to that of light \cite{TheLIGOScientific:2017qsa,Monitor:2017mdv}. This reduces the number of redshift-dependent functions in the effective description that govern how modifications of gravity affects perturbations to three:
\bea
\am(z) &-& \mbox{evolution of the effective Planck mass}; \notag\\
\ab(z) &-& \mbox{mixing between the metric and the DE field};\notag\\
\ak(z) &-& \mbox{kinetic energy of scalar perturbations}. \notag
\eea

Although $\ak$ has virtually no effect on constraints from current data \cite{Bellini:2015xja,Frusciante:2018jzw}, it needs to be included as a free parameter, since it regulates the propagation speed of DE perturbations. Setting it arbitrarily to zero could restrict the space of stable models and thus bias the constraints \cite{Kreisch:2017uet,Frusciante:2018jzw}.

The functional forms of $\alpha_I(z)$, $I= M,B,K$, are not given by the effective description. For simplicity, we use the effective DE parametrisation\cite{Piazza:2013pua,Bellini:2014fua} common in literature:
\bea
\alpha_I(z) = a_I\,\frac{ \Omega_{\rm x}(z)}{\Omega_{\rm x,0}}.
\eea

We also allow for deviations from a $\Lambda$CDM background by using the Chevallier-Polarski-Linder (CPL) \cite{Chevallier:2000qy,Linder:2002et} parametrisation for the effective dark energy (DE) equation of state of the Horndeski models:
\bea
w_{\rm x}(z) = w_0 + w_a \,\frac{z}{1+z}\,.
\eea

In summary, the Horndeki model we consider contains five additional free parameters with respect to $\Lambda$CDM
\begin{equation}
\big\lbrace \omb,\omc, H_0, \tau,A_{\rm s}, n_{\rm s},\sumnu, w_0,w_a, a_M, a_B,a_K \big\rbrace .
\end{equation}
$\Lambda$CDM is recovered for $w_0 = -1$ and $w_a = a_M = a_B = a_K = 0$.

\section{Methodology}\label{sec:Method}

The cosmological evolution of the models is computed using the Boltzmann code\footnote{\url{www.class-code.net}} \texttt{CLASS} \cite{Blas:2011rf}, and its modified version\footnote{\url{www.hiclass-code.net}} \texttt{hi\_class} \cite{Zumalacarregui:2016pph,Bellini:2019syt}. The cosmological data -- hereafter referred to as the ``baseline" -- contains the SDSS-II/SNLS3 Joint Light-curve Analysis (JLA) sample of SNIa \cite{Betoule:2014frx}, the BOSS baryon acoustic oscillation (BAO) measurements \cite{Beutler:2011hx,Anderson:2013zyy,Ross:2014qpa} and the Planck 2018 cosmic microwave background (CMB) data, the low- and high-multipole temperature and polarisation \cite{Aghanim:2019ame}. We choose not to include CMB lensing data, to avoid inconsistencies related to potential $\Lambda$CDM-dependent assumptions made during the lensing reconstruction.

Our aim is to focus on the gain from breaking growth degeneracy, rather than making realistic mocks and forecasts. In order to compare the constraining power of separated measurements of $f$ and $\sig$ with the combined measurements $f\sig$, we simulate data for a nominal future galaxy sample that delivers a one percent precision for $f$, $\sig$ and $\fsig$. We assume a redshift range containing 10 measurements at $z = 0.1, 0.2, ..., 1$. The effects of extending the redshift range are studied in Section \ref{sec:extredshift}. We anticipate that a Stage IV experiment conducting a spectroscopic galaxy count survey together with a weak lensing survey, such as Euclid \cite{Amendola:2016saw}, should be able to achieve close to 1\% precision on $\fsig$, $f$ and $\sig$, using Planck priors on standard cosmological parameters.

Whenever needed, the growth quantities are computed with \texttt{CLASS} or \texttt{hi\_class}. In order to compare the constraints on the same footing and avoid non-linear model dependencies, we compute the growth quantities with the linear power spectrum only. The values of $\sig$ are obtained via the usual weighted integral of the linear power spectrum and $f$ is computed as the log derivative $f=-(1+z)d \ln \sig/ d \ln z$ for simplicity.

We use as fiducial parameters the best-fit values obtained from the baseline constraints for the $\Lambda$CDM and Horndeski models. Then we create three sets of mocks for both models (for $f$, $\sig$ and $\fsig$), each exactly centred on their fiducial, $i.e.$ with no random variance added to the data. $\Lambda$CDM has been shown to lie in a corner of the parameter space of stable Horndeski models \cite{Piazza:2013pua}, i.e., ghost- and gradient-free models. When performing forecasts using Markov chain Monte Carlo (MCMC) methods, the stability priors can lead to a disfavouring of models lying close to the corner, purely due to volume effects and independently of their actual likelihood. Such considerations may have a significant effect on our results. This is hinted at for example by the highly irregular posteriors in the baseline case in Figure \ref{fig:horn_triangle} (grey contours) and the mismatch between their maximum and the best-fit model (dotted lines), characteristic of non-negligible prior effects. We can however expect those effects to be mitigated when additional data is added to the analysis, due to the fact that our Horndeski fiducial model (derived from the baseline best-fit and used to produce our mocks) lies noticeably away from the  ``$\Lambda$CDM corner". Even if our MCMC explorations were impacted by such priors, this should not affect our conclusions since we always make statements regarding relative improvements.

\section{Constraints}

The sampling of all the considered likelihoods, as well as the computation of best-fit parameters, are performed using the publicly available\footnote{\url{https://github.com/s-ilic/ECLAIR}} suite of codes ECLAIR \cite{Ilic:2020onu}. It uses as its main sampling algorithm the affine-invariant ensemble method of \cite{ForemanMackey:2012ig} and contains a novel and robust maximiser with reliable convergence towards the global maximum of the posterior.

\subsection{$\Lambda$CDM}\label{LCDM-constr}

\begin{figure*}[!]
\begin{center}
\includegraphics[scale=0.36]{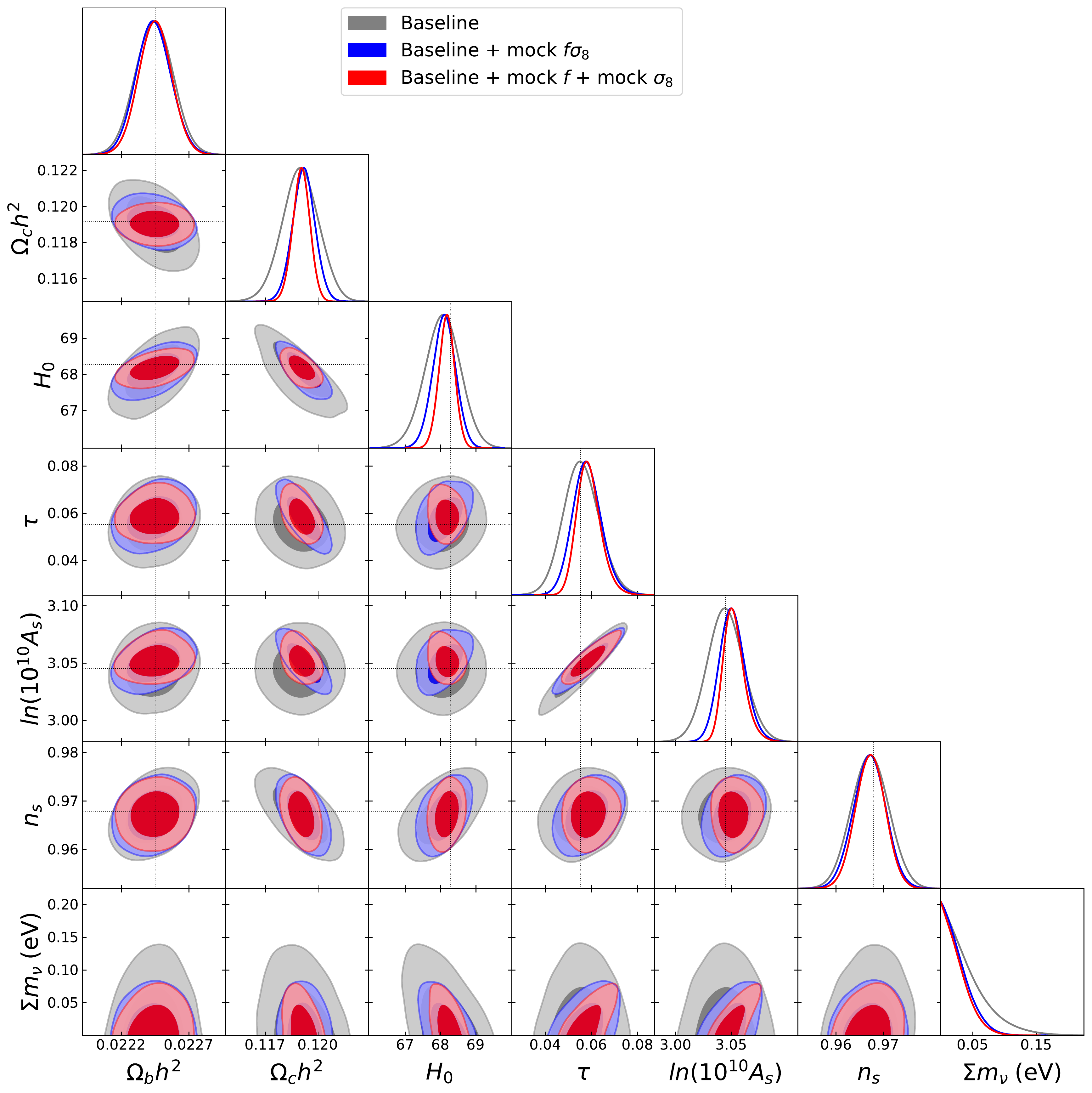}
\caption{1D and 2D marginalised posterior distributions for $\Lambda$CDM parameters derived from the baseline only (grey), baseline with mock on $\fsig$ (blue) and baseline with mocks on $f$ and $\sig$ (red). The dotted lines indicate the parameter values for the fiducial model (corresponding to the baseline best-fit) used when generating mocks.}
\label{fig:lcdm_triangle}
\end{center}
\end{figure*}

Marginalised posterior distributions are shown in Figure \ref{fig:lcdm_triangle}. The corresponding means and 68\% confidence intervals are given in Table \ref{tab:lcdm_cons}, while Table \ref{tab:lcdm_imp} shows the gain in precision relative to baseline (first two columns) and for the separated growth measurements $f+\sig$ relative to the standard $f\sig$ measurements (last column). We define the precision as the inverse width of the 68\% marginalised confidence interval rather than using relative errors, since the latter can become misleading when the mean values are close to zero (e.g., in the case of $\sumnu$). In addition, comparing relative errors would also be biased when the mean values shift, as happens for the Horndeski models (see below).

\renewcommand{\arraystretch}{1.8}
\begin{table}[!]
\begin{center}
\begin{tabular}{|c|c|c|}
\hline
                        & $\bm \fsig$ & $\bm f$ + $\bm \sig$ \\
\hline
$\bm \omb$              & $ 0.02244^{0.00013}_{-0.00013} $ &	$ 0.02245^{0.00012}_{-0.00012} $ \\
\hline
$\bm \omc$              & $ 0.11918^{0.00062}_{-0.00063} $ &	$ 0.11904^{0.00048}_{-0.00048} $ \\
\hline
$\bm{H_0}$              & $ 68.10^{0.32}_{-0.32} $ &	$ 68.17^{0.22}_{-0.22} $ \\
\hline
$\bm \tau$              & $ 0.0581^{0.0058}_{-0.0066} $ &	$ 0.0589^{0.0044}_{-0.0056} $ \\
\hline
$\bm{\ln (10^{10}A_{\rm s})}$ & $ 3.0509^{0.0106}_{-0.0121} $ &	$ 3.0524^{0.0075}_{-0.0104} $ \\
\hline
$\bm{n_{\rm s}}$              & $ 0.9671^{0.0034}_{-0.0034} $ &	$ 0.9673^{0.0031}_{-0.0031} $ \\
\hline
$\bm \sumnu$            & $ 0.0263^{0.0062}_{-0.0263} $ &	$ 0.0248^{0.0058}_{-0.0248} $ \\
\hline
\end{tabular}
\caption{Mean and 68\% confidence interval for $\Lambda$CDM parameters. The constraints are obtained by combining the baseline with the $f\sig$ mock (middle column) and $f$ and $\sig$ mocks (right column).}
\label{tab:lcdm_cons}
\end{center}
\end{table}

Next-generation surveys are forecast to deliver improved constraints from high-precision RSD $f\sig$  data (see e.g. \cite{Amendola:2016saw,Bacon:2018dui}). The triangle plots and the tables confirm this. Table \ref{tab:lcdm_imp} (first column) shows that the gain in precision ranges from $\sim\! 10\%$ for $\omb$ up to more than $\sim\! 50\%$ for $ \omc, H_0$ and $\sumnu$, when considering the addition of the mock data on $\fsig$ with 1\% relative error to current cosmological datasets. 

As expected the constraints improve further with the split mock data on $f$ and $\sig$, each with a 1\% relative error. This combination performs from 6\% to almost 50\% better. In particular, the precision on $\omc$ and $H_0$ is more than doubled relative to the baseline data alone.

\renewcommand{\arraystretch}{1.3}
\begin{table}[!]
\begin{center}
\begin{tabular}{|c|c|c|c|}
\hline
& \bf{baseline}  & \bf{baseline} &\bf{baseline}   \\
& \bf{+} $\bm{\fsig}$   & \bf{+} $\bm f$ + $\bm{\sig}$ & \bf{+} $\bm{f}$ + $\bm{\sig}$   \\
                        & \bf{/ baseline}                & \bf{/ baseline}               & \bf{/ baseline +} $\bm{\fsig}$  \\
\hline
$\bm \omb$              & $ 1.08 $ & $ 1.15 $ & $ 1.06 $ \\
\hline
$\bm \omc$              & $ 1.66 $ & $ 2.16 $ & $ 1.30 $ \\
\hline
$\bm{H_0}$              & $ 1.55 $ & $ 2.26 $ & $ 1.46 $ \\
\hline
$\bm \tau$              & $ 1.25 $ & $ 1.54 $ & $ 1.22 $ \\
\hline
$\bm{\ln (10^{10}A_{\rm s})}$ & $ 1.40 $ & $ 1.77 $ & $ 1.26 $ \\
\hline
$\bm{n_{\rm s}}$              & $ 1.16 $ & $ 1.27 $ & $ 1.09 $ \\
\hline
$\bm \sumnu$            & $ 1.48 $ & $ 1.57 $ & $ 1.13 $ \\
\hline
\end{tabular}
\caption{Precision ratios for $\Lambda$CDM parameters. See Section \ref{LCDM-constr} for details.}
\label{tab:lcdm_imp}
\end{center}
\end{table}

The improvement obtained from the split $f$ and $\sig$ data over $\fsig$ (as quantified by the third column of Table \ref{tab:lcdm_imp}) does not lead to an equal increase in precision on all the parameters that were already well constrained with $\fsig$ RSD data. As an example, we can compare $\sumnu$ and $H_0$. Adding $f\sig$ data yields almost a 50\% gain on $\sumnu$, while the split $f+\sig$ data further increases the precision by 13\%. By contrast, $H_0$ precision first increases by 55\% followed by another 46\% with the splitting.

\begin{figure*}[!]
\begin{center}
\hspace{7.4mm}\includegraphics[clip, trim = 0cm 1.0cm 0cm 1.2cm, width=11.86cm]{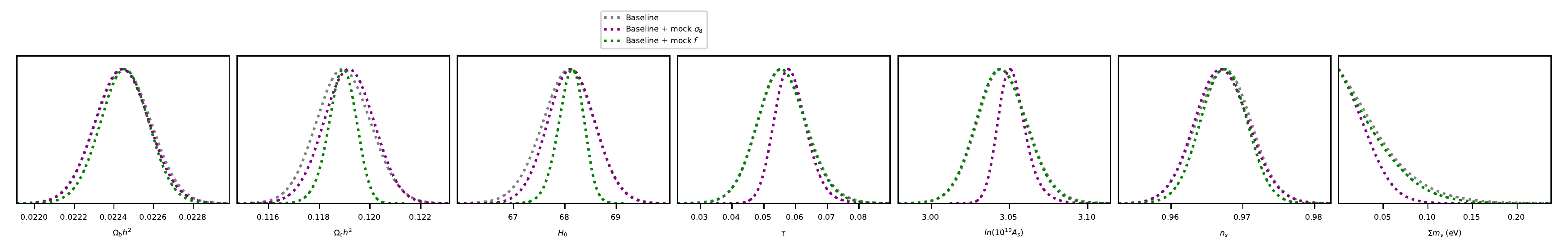}
\includegraphics[clip, trim = 0cm 0.0cm 0cm 0.7cm, width=12.5cm]{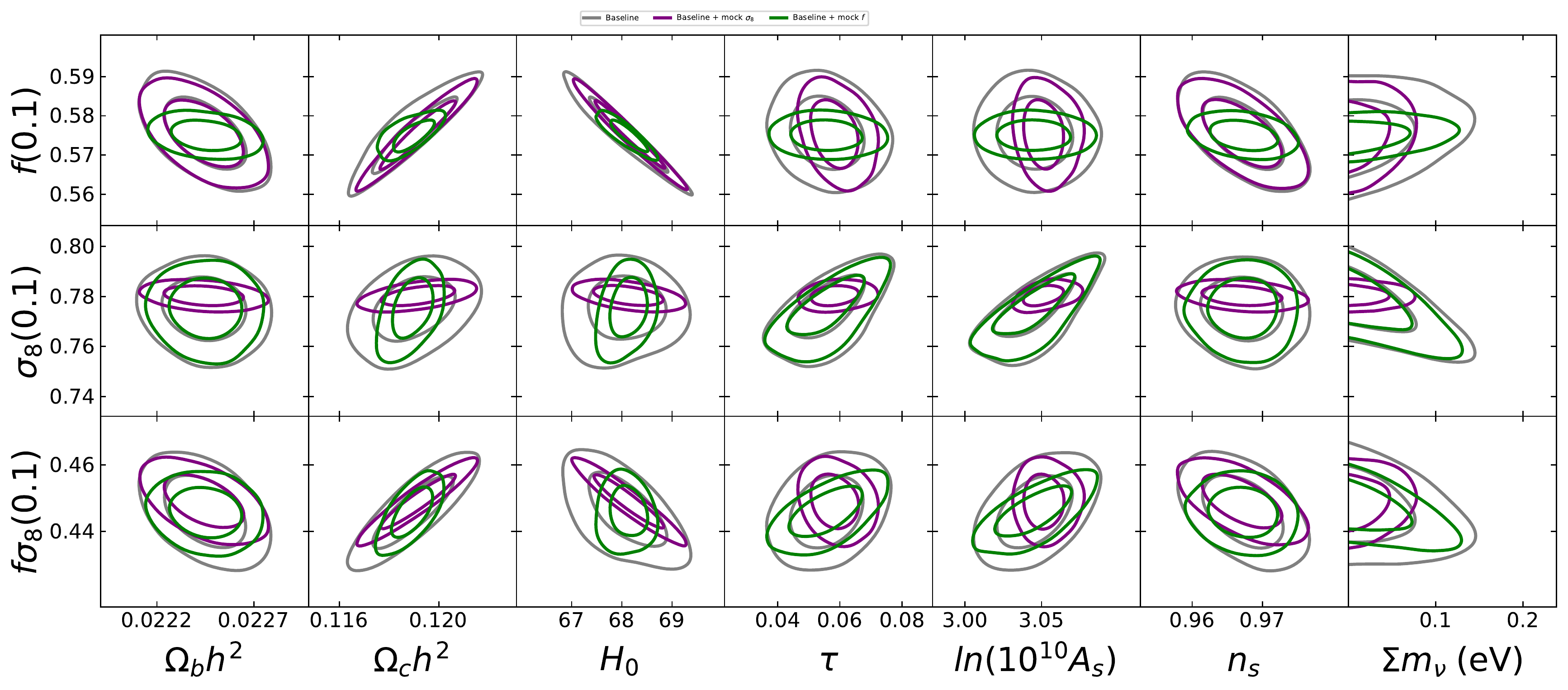}
\includegraphics[clip, trim = 17cm 13.7cm 17cm 0cm, width=10cm]{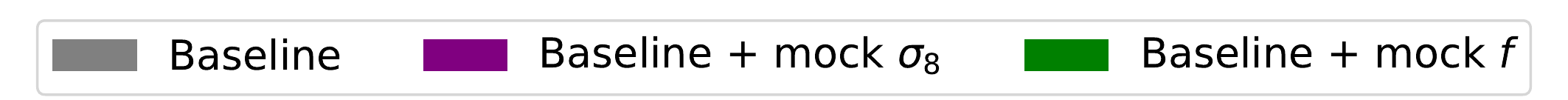}
\caption{1D marginalised posterior distributions (top row) for $\Lambda$CDM parameters, from baseline only (grey), baseline + mock on $f$ (green) and baseline + mock on $\sig$ (purple). Rows below show 2D posteriors of cosmological parameters against derived parameters $f$, $\sig$ and $\fsig$ at $z = 0.1$.}
\label{fig:lcdm_rectangle}
\end{center}
\end{figure*}

The growth probes $f$, $\sig$, and $\fsig$ have different sensitivities to each cosmological parameter, which explains the range of changes in precision. One way to examine those sensitivities is to start with the baseline-only constraints. Figure \ref{fig:lcdm_rectangle} shows the posterior distributions of $f$, $\sig$ and $\fsig$ at redshift $z=0.1$ as derived parameters versus the cosmological parameters\footnote{We find the orientations of these posteriors (i.e., correlation factors between parameters) to change very little with redshift. Therefore we consider only $z=0.1$ for illustration, but our discussion applies to the other $z$.}. Each posterior thus illustrates how a change in a given cosmological parameter impacts the values of the derived growth quantities, taking into account (i.e., marginalising over) the remaining cosmological parameters and how their values need to change to keep a decent fit to the data. 

On the other hand, adding constraints on the growth quantities amounts to convolving their posteriors with a Gaussian distribution (with a width equal to 1\% of the central value). This in turn may reduce the width of the posterior on cosmological parameters, depending on the amount of correlation between the two. It is thus expected that cosmological parameters that are highly correlated (i.e., thin tilted ellipses) with a given growth quantity in the baseline case, will show the best improvements after including measurements of that growth quantity.

From Figure \ref{fig:lcdm_rectangle} we find that $\Omega_{\rm b}$, $\Omega_{\rm c}$, $H_0$, $n_{\rm s}$ are better constrained by adding the $f$ mock (green) to the baseline, while $\tau$, $A_{\rm s}$, $\sumnu$ are better constrained by adding the $\sig$ mock (purple). This may appear counter to the common expectation that $\sig$ is more sensitive to parameters affecting the power spectrum amplitude, while $f$ is more sensitive to parameters affecting its shape. It is the correlations induced by the baseline constraints that are the decisive factor.

Let us consider an illustrative example from Figure \ref{fig:lcdm_rectangle}: the 2D posterior of $\lbrace f(0.1), H_0\rbrace$ exhibits a high correlation  (thin tilted ellipse), while that of $\lbrace \sig (0.1), H_0\rbrace$ is relatively irregular and close to an uncorrelated case. As a result, the addition of the $f$ mock improves the $H_0$ constraint significantly more relative to the baseline (see the 1D posterior of $H_0$ in the top row of Figure \ref{fig:lcdm_rectangle}).

These correlations can even lead to improved constraints on parameters that $f$ and $\sig$ should not depend on. An example is the tight constraint on the reionisation parameter $\tau$ produced by the mock on $\sig$, which originates in the tight constraint on $A_{\rm s}$ from $\sig$, combined with the underlying high correlation between $A_{\rm s}$ and $\tau$, as shown in Figure \ref{fig:lcdm_triangle}. A tight constraint on $\tau$ is obtained even though it does not play a role in the value of $\sig$.

\subsection{Horndeski}\label{HORN-constr}

\begin{figure*}[!]
\begin{center}
\includegraphics[scale=0.31]{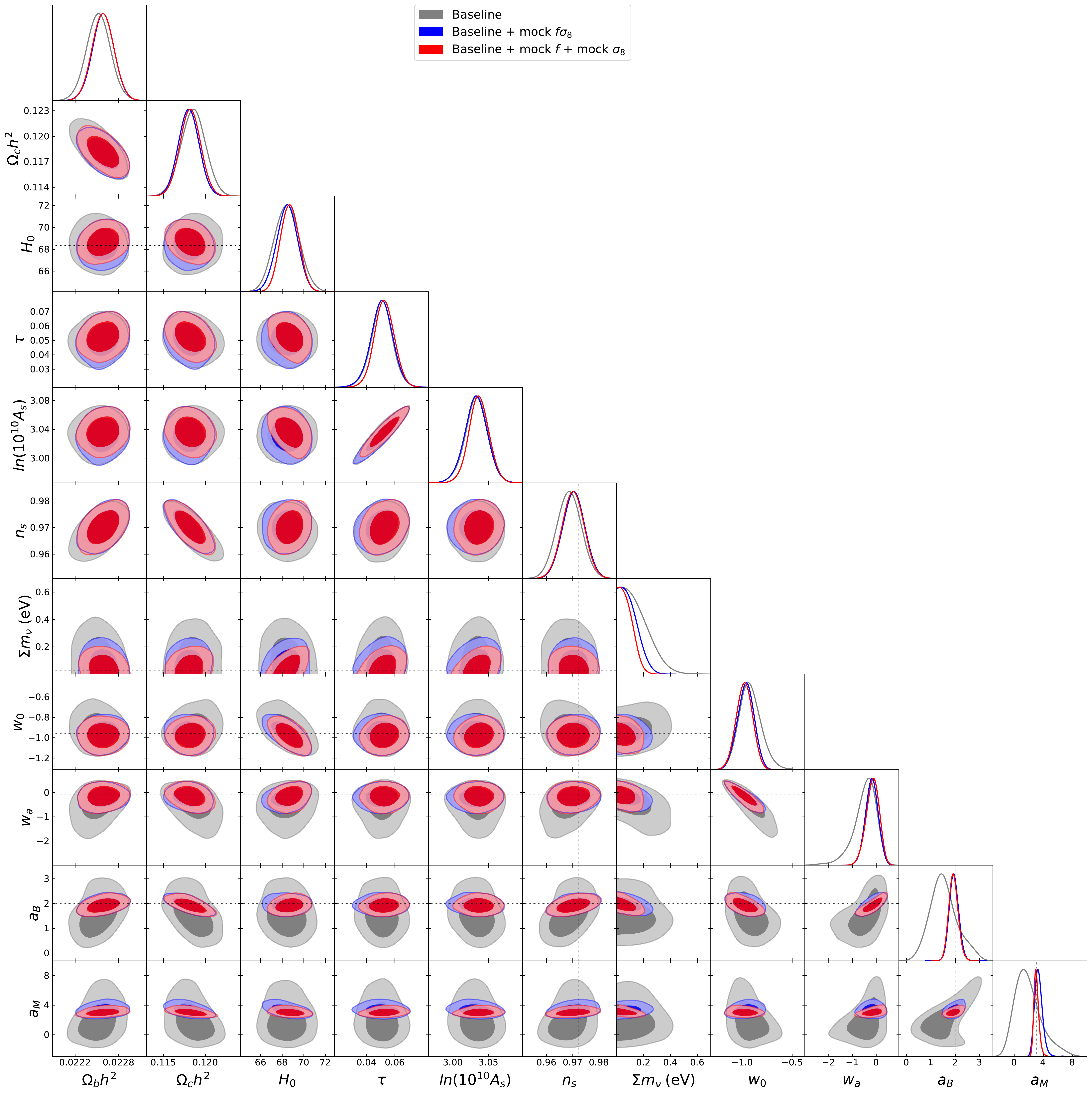}
\caption{1D and 2D marginalised posterior distributions for Horndeski parameters derived from the baseline only (grey), baseline with mock on $\fsig$ (blue) and baseline with mocks on $f$ and $\sig$ (red). The dotted lines indicate the parameter values for the fiducial model (corresponding to the baseline best-fit) used when generating mocks.}
\label{fig:horn_triangle}
\end{center}
\end{figure*}

The Horndeski parameter space is extended to include modifications in the background ($w_0,w_a$) and in the perturbations ($\am,\ak,\ab$). Marginalised posterior distributions with the baseline and mock data sets are displayed in Figure \ref{fig:horn_triangle}, with the corresponding means and 68\% confidence intervals in Table \ref{tab:horn_cons}. We observe that the maximum of the posterior distribution for the extension parameters shifts significantly towards the best-fit model (dotted lines), while the contours assume a much more regular, ellipsoidal shape compared to the baseline case. This is expected in a transition from a regime where priors still play a significant role (as discussed at the end of Section \ref{sec:Method}), to a situation where data dominate the posterior.

Interestingly, these results also show that if the true underlying cosmology is indeed close to the Horndeski best-fit fiducial, then growth data with 1\% relative precision (over the redshift range considered) could lead to the detection of this deviation from $\Lambda$CDM with strong significance (more than 5$\sigma$).

\renewcommand{\arraystretch}{1.8}
\begin{table}[!]
\begin{center}
\begin{tabular}{|c|c|c|c|c|}
\hline
 & $\bm \fsig$ & $\bm f$ + $\bm \sig$ \\
\hline
$\bm \omb$              & $ 0.02259^{0.00015}_{-0.00014} $ &	$ 0.02258^{0.00015}_{-0.00015} $ \\
\hline
$\bm \omc$              & $ 0.11801^{0.00122}_{-0.00122} $ &	$ 0.11819^{0.00122}_{-0.00122} $ \\
\hline
$\bm{H_0}$              & $ 68.44^{0.96}_{-0.96} $ &	$ 68.69^{0.85}_{-0.85} $ \\
\hline
$\bm \tau$              & $ 0.0508^{0.0075}_{-0.0075} $ &	$ 0.0526^{0.0070}_{-0.0068} $ \\
\hline
$\bm{\ln (10^{10}A_{\rm s})}$ & $ 3.0324^{0.0157}_{-0.0155} $ &	$ 3.0366^{0.0140}_{-0.0136} $   \\
\hline
$\bm{n_{\rm s}}$              & $ 0.9704^{0.0042}_{-0.0042} $ &	$ 0.9702^{0.0042}_{-0.0042} $ \\
\hline
$\bm \sumnu$            & $ 0.0953^{0.0261}_{-0.0953} $ &	$ 0.0742^{0.0206}_{-0.0742} $  \\
\hline
$\bm{w_0}$              & $ -0.9636^{0.0862}_{-0.0797} $ &	$ -0.9770^{0.0813}_{-0.0814} $   \\
\hline
$\bm{w_a}$              & $ -0.1901^{0.2632}_{-0.2636} $ &	$ -0.1543^{0.2984}_{-0.2494} $  \\
\hline
$\bm{a_B}$              & $ 1.9493^{0.1801}_{-0.2058} $ &	$ 1.9262^{0.1791}_{-0.2003} $  \\
\hline
$\bm{a_M}$              & $ 3.3473^{0.4411}_{-0.5943} $ &	$ 3.0485^{0.2630}_{-0.3799} $ \\
\hline
\end{tabular}
\caption{Mean and 68\% confidence interval for Horndeski parameters. The constraints are obtained by combining the baseline with the $f\sig$ mock (middle column) and $f$ and $\sig$ mocks (right column).}
\label{tab:horn_cons}
\end{center}
\end{table}

Table \ref{tab:horn_imp} shows the gain in precision relative to baseline (first two columns) and for the separated growth measurements $f+\sig$ relative to the standard $f\sig$ measurements (last column). As pointed out earlier, the kineticity coupling $\ak$ is not constrained by the data and is therefore not included in the figure and tables. but $a_K$ is included as a free parameter in the analysis. The accuracy that was gained on the cosmological parameters in $\Lambda$CDM is largely lost. Adding the mock on $\fsig$ only delivers up to $\sim\!20\%$ precision gain (see Table \ref{tab:horn_imp}). This can be attributed to the addition of new, poorly constrained degrees of freedom which naturally leads to larger errors on all the original parameters via correlations, as both sets may have similar and degenerate effects on the growth of structure. For example, Figure \ref{fig:horn_triangle} shows how $a_M$ and $a_B$ are relatively degenerate with other parameters when using the baseline data only.

However, there is significant improvement for the extension parameters: adding future $\fsig$ data yields a 230\% improvement for the running of the effective Planck mass $\am$ and a remarkable $\sim\!50\%$ gain for $\sumnu$. Even though $\fsig$ is a probe of the perturbations, adding its mock to the baseline achieves a surprising $\sim\!30\%$ and $\sim\!60\%$ gain in precision for $w_0$ and $w_a$ respectively.

The additional gain from disentangling $f$ and $\sig$ measurements is also subject to the effects of opening up the parameter space. The standard parameters see little improvement ($<15\%$) over the $\fsig$ case. By contrast, $w_a,\sumnu$ and $\am$ precisions jump by a further $\sim\!20\%, \sim\!65\%$ and $\sim\!80\%$ respectively.

\renewcommand{\arraystretch}{1.34}
\begin{table}[!]
\begin{center}

\begin{tabular}{|c|c|c|c|}
\hline
& \bf{baseline}  & \bf{baseline} &\bf{baseline}   \\
& \bf{+} $\bm{\fsig}$   & \bf{+} $\bm f$ + $\bm{\sig}$ & \bf{+} $\bm{f}$ + $\bm{\sig}$   \\
                        & \bf{/ baseline}                & \bf{/ baseline}               & \bf{/ baseline +} $\bm{\fsig}$  \\
\hline
$\bm \omb$              & $ 1.11 $ & $ 1.10 $ & $ 0.99 $\\
\hline
$\bm \omc$              & $ 1.20 $ & $ 1.20 $ & $ 1.00 $ \\
\hline
$\bm{H_0}$              & $ 1.19 $ & $ 1.35 $ & $ 1.12 $ \\
\hline
$\bm \tau$              & $ 0.96 $ & $ 1.04 $ & $ 1.05 $ \\
\hline
$\bm{\ln (10^{10}A_{\rm s})}$ & $ 0.98 $ & $ 1.11 $ & $ 1.13 $ \\
\hline
$\bm{n_{\rm s}}$              & $ 1.10 $ & $ 1.10 $ & $ 1.00 $ \\
\hline
$\bm \sumnu$            & $ 1.49 $ & $ 1.91 $ & $ 1.65 $ \\
\hline
$\bm{w_0}$              & $ 1.29 $ & $ 1.32 $ & $ 1.01 $ \\
\hline
$\bm{w_a}$              & $ 1.65 $ & $ 1.59 $ & $ 1.18 $ \\
\hline
$\bm{a_B}$              & $ 2.83 $ & $ 2.87 $ & $ 1.03 $ \\
\hline
$\bm{a_M}$              & $ 3.30 $ & $ 5.32 $ & $ 1.77 $ \\
\hline
\end{tabular}
\caption{Precision ratios for Horndeski parameters. See Section \ref{HORN-constr} for details.}
\label{tab:horn_imp}
\end{center}
\end{table}

The underlying reason that growth data provide such an enhancement on precision for the Horndeski parameters is rooted in the modification of gravitational dynamics (e.g., the Poisson equation) by $\alpha_I$. As discussed in \cite{Perenon:2019dpc}, these modifications produce two opposing contributions:\\
\textasteriskcentered~  a fifth force, enhancing growth;\\
\textasteriskcentered~  a higher effective Planck mass, suppressing growth.

The effective Planck mass is controlled solely by $\am$ for the models we consider. As a result, growth data strongly constrains $a_M$ and also $a_B$. Table \ref{tab:horn_imp} shows that the splitting of $\fsig$ into $f$ and $\sig$ is very effective to further constrain $a_M$, thereby disentangling the fifth force and effective Planck mass contributions. This feature was seen even with current split data in \cite{Perenon:2019dpc}.

\begin{figure*}[!]
\begin{center}
\hspace{6.5mm}\includegraphics[clip, trim = 0cm 1.0cm 0cm 1.2cm, width=16.95cm]{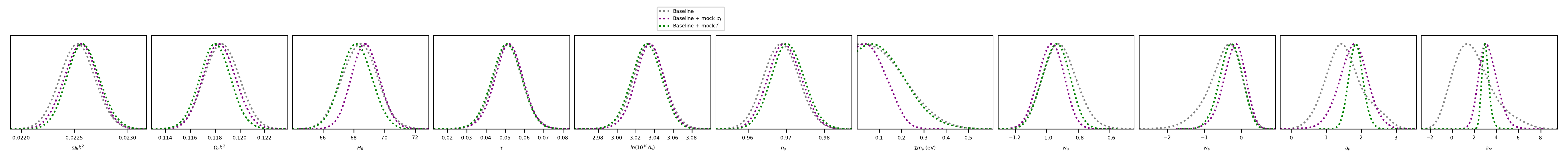}
\includegraphics[clip, trim = 0cm 0.0cm 0cm 0.7cm, width=17.5cm]{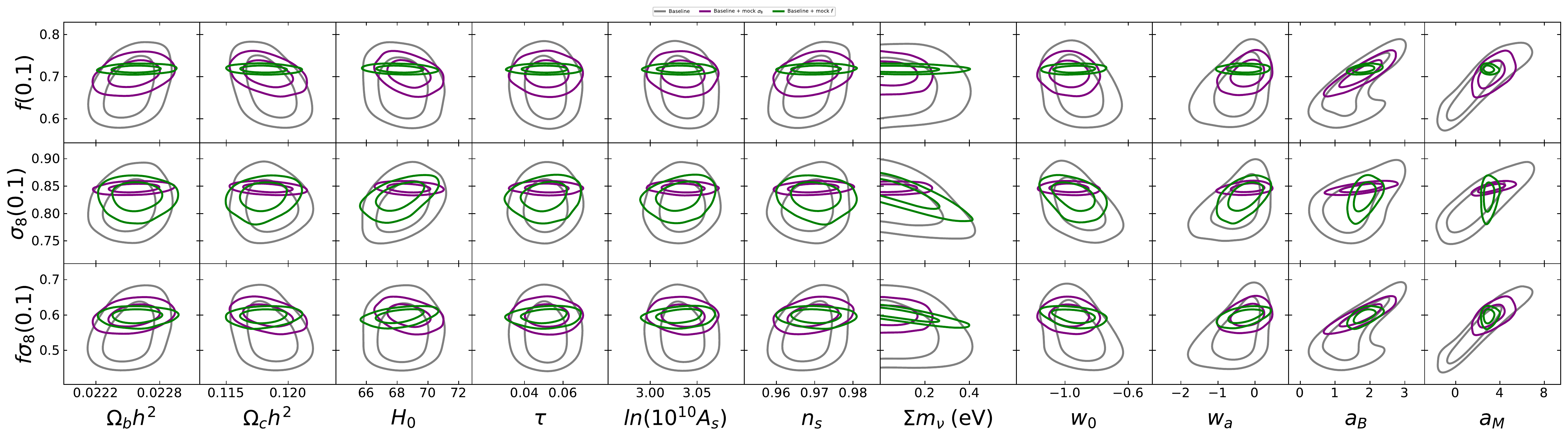}
\hspace{10mm}\includegraphics[clip, trim = 17cm 13.7cm 17cm 0cm, width=10cm]{PLOT_rectangle_legend}
\caption{1D marginalised posterior distributions (top row) for Horndeski parameters, from baseline only (grey), baseline + mock on $f$ (green) and baseline + mock on $\sig$ (purple). Rows below show 2D posteriors of cosmological parameters against derived parameters $f$, $\sig$ and $\fsig$ computed at $z = 0.1$.}
\label{fig:horn_rectangle}
\end{center}
\end{figure*}

The modified background parameters $w_0,\,w_a$ contribute also to the growth of structure through the Hubble friction. Their effects on growth are therefore degenerate with those of $\alpha_I$. We see in Figure \ref{fig:horn_rectangle} that $ w_0,\,w_a,\,a_B,\,a_M$ display some degeneracies in their 2D marginalised posteriors.

Following the arguments for $\Lambda$CDM, we can understand the separate improvements from $f$ and $\sig$ by analysing their posterior distributions versus cosmological parameters, shown in Figure \ref{fig:horn_rectangle}. Note that the stability requirements for the Horndeski models induce highly non-Gaussian posterior distributions, which makes the analysis more subtle. Figure \ref{fig:horn_rectangle} shows that $f$ correlates more strongly with $a_B,\, a_M$ than $\sig$, so that adding $f$ measurements results in a larger increase in precision on these parameters. Since these two parameters control the strength of the fifth force, this could be expected, given that $\sig$ is an integrated function of $f$, which tends to wash out the effects of the fifth force. Note that the fifth force is an effect occurring at low redshifts as opposed to the effect of the Hubble friction or neutrinos. A chain of correlations -- seen in the baseline constraints -- shows that $\sig$ brings a larger gain in precision for $w_0,\,w_a$ and $\sumnu$. This signals therefore a higher sensitivity of $\sig$ to modifications of gravity spanning longer periods.

It is in fact expected that the effect of neutrinos is partially degenerate with that of modified gravity (see e.g. \cite{Wright:2019qhf,Ballardini:2020iws}). Massive neutrinos suppress the growth of structure on small scales, which can either oppose or reinforce modified gravity, depending on whether the fifth force or the Planck mass running is favoured. Horndeski models compatible with current RSD $\fsig$ constraints produce a suppression of growth at late times \cite{Perenon:2019dpc}. 

The baseline constraints in Figure \ref{fig:horn_triangle} show that the 2D posteriors of $\sumnu$ with $a_B$ and $a_M$ are fairly irregularly shaped , while those with $w_0$ and $w_a$ are more correlated. More surprisingly, as noted above, $\sumnu$ has almost a 50\% gain with the addition of the $\fsig$ mock data, as in the case of $\Lambda$CDM. The splitting improves constraints by a further 70\% as opposed to 7\% in $\Lambda$CDM. It is therefore clear that these growth mocks break the neutrino-modified gravity degeneracy by constraining efficiently $\sumnu$ and the extension parameters. Figure \ref{fig:horn_rectangle} tells us that this is rooted in the correlation of $\sumnu$ with $\sig$ in the baseline. 

On the other hand, we also see that all the intricate degeneracies between the extension parameters and standard model parameters render the baseline constraints for the latter much less correlated than in the case of $\Lambda$CDM. This explains why the improvements from the splitting are not as great in the case of Horndeski for the other standard parameters.

Note that when the background evolution is fixed to that of $\Lambda$CDM, $\sumnu$ displays correlation with $\ab$ \cite{Bellomo:2016xhl}. Here, the freedom that arises from varying $w_0,\, w_a$ lessens that correlation.

\subsection{Extending the redshift range}\label{sec:extredshift}

Having understood better the influence of each mock data set on the constraints, we now assess the effect of extending the redshift coverage of the mocks. More specifically, we examine the respective merits of adding $\fsig$ or $f+\sig$ measurements, when extending the maximum redshift of each mock. Table \ref{tab:extredshift_cons} shows that the combined data $\fsig$ with $z_{\rm max}=2$ (first column) performs no better than $f$ + $\sig$ data with half the redshift range ($z_{\rm max}=1$, see Tables \ref{tab:lcdm_cons} and \ref{tab:horn_cons}). We find that extending the redshift range further improves the precision up to 30\% with respect to $z_{\rm max}=1$ in the case of the combined mock $\fsig$ for $\Lambda$CDM and Horndeski models, and respectively 20\% and 15\% in the case of $f$ and $\sig$ mocks.

\renewcommand{\arraystretch}{1.8}
\begin{table}[!]
\begin{center}
\begin{tabular}{|c|c|c|c|c|}
\hline
                        & $\bm{\fsig}$ $(z_{\rm max}=2)$ & $\bm{f}$ + $\bm{\sig}$ $(z_{\rm max}=2)$  \\
\hline
$\bm{\omb}$             & $ 0.02245^{0.00012}_{-0.00012} $ &	$ 0.02244^{0.00012}_{-0.00012} $ \\
\hline
$\bm{\omc}$             & $ 0.11910^{0.00055}_{-0.00054} $ &	$ 0.11907^{0.00046}_{-0.00046} $ \\
\hline
$\bm{H_0}$              & $ 68.17^{0.25}_{-0.25} $ &	$ 68.17^{0.21}_{-0.21} $  \\
\hline
$\bm{\tau}$             & $ 0.0588^{0.0045}_{-0.0056} $ &	$ 0.0587^{0.0038}_{-0.0048} $   \\
\hline
$\bm{\ln (10^{10}A_{\rm s})}$ & $ 3.0524^{0.0075}_{-0.0102} $ &	$ 3.0521^{0.0062}_{-0.0089} $ \\
\hline
$\bm{n_{\rm s}}$              & $ 0.9673^{0.0032}_{-0.0033} $ &	$ 0.9673^{0.0031}_{-0.0031} $ \\
\hline
$\bm{\sumnu}$           & $ 0.0234^{0.0055}_{-0.0234} $ &	$ 0.0234^{0.0055}_{-0.0234} $ \\
\hline
\hline
$\bm{\omb}$             & $ 0.02259^{0.00015}_{-0.00015} $ &	$ 0.02258^{0.00015}_{-0.00015} $ \\
\hline
$\bm{\omc}$             & $ 0.11813^{0.00123}_{-0.00123} $ &	$ 0.11824^{0.00117}_{-0.00117} $  \\
\hline
$\bm{H_0}$              & $ 68.60^{0.88}_{-0.88} $ &	$ 68.63^{0.82}_{-0.81} $ \\
\hline
$\bm{\tau}$             & $ 0.0523^{0.0071}_{-0.0072} $ &	$ 0.0524^{0.0071}_{-0.0071} $ \\
\hline
$\bm{\ln (10^{10}A_{\rm s})}$ & $ 3.0357^{0.0144}_{-0.0143} $ &	$ 3.0365^{0.0141}_{-0.0143} $ \\
\hline
$\bm{n_{\rm s}}$              & $ 0.9702^{0.0042}_{-0.0042} $ &	$ 0.9701^{0.0041}_{-0.0041} $ \\
\hline
$\bm{\sumnu}$           & $ 0.0656^{0.0181}_{-0.0656} $ &	$ 0.0693^{0.0185}_{-0.0693} $ \\
\hline
$\bm{w_0}$              & $ -0.9761^{0.0840}_{-0.0850} $ &	$ -0.9764^{0.0705}_{-0.0710} $ \\
\hline
$\bm{w_a}$              & $ -0.1328^{0.2831}_{-0.2542} $ &	$ -0.1452^{0.2666}_{-0.2167} $ \\
\hline
$\bm{a_B}$              & $ 1.9494^{0.1710}_{-0.1939} $ &	$ 1.9266^{0.1594}_{-0.1857} $ \\
\hline
$\bm{a_M}$              & $ 3.1274^{0.3142}_{-0.4618} $ &	$ 3.0552^{0.2409}_{-0.3672} $	\\
\hline
\end{tabular}
\caption{Mean and 68\% confidence interval for $\Lambda$CDM (top) and Horndeski (bottom) parameters with the redshift of the mocks extended to $z = 0.1,\, 0.2,\, ...,\, 2.0$. The constraints are obtained by combining the baseline with the $f\sig$ mock (middle column) and $f$ and $\sig$ mocks (right column).}
\label{tab:extredshift_cons}
\end{center}
\end{table}

\section{Conclusion}

Upcoming galaxy surveys such as Euclid \cite{Amendola:2016saw} and SKA \cite{Bacon:2018dui} with their unprecedented precision is a call to sharpen our tools for  constraining gravity. One cosmological probe well-suited for that task is the growth of structure. This toolbox is further complemented by the releases of measurements on $f$ and $\sig$ \cite{delaTorre:2016rxm,Shi:2017qpr,Jullo:2019lgq,Gil-Marin:2016wya}.

In this paper, we considered the performance that a future nominal galaxy sample can deliver with a $\sim 1\%$ relative error on $f$ and $\sig$ separately and on the combination $f\sig$. We compared the constraints from the separated data with those from the combination data. We assumed 10 measurements per growth quantity equally spread on the redshift range $z = 0.1,\, 0.2,\, ...,\, 1.0$. For the case of $\Lambda$CDM, the improvements in precision range over $\sim\! 5$--$50\%$. For modified gravity described by Horndeski models, the improvements on these standard model parameters reduce to $\sim\! 0$--$15\%$. 

However, the splitting of $f$ and $\sig$ stands out as very effective in breaking the neutrino - modified gravity degeneracy, with the sum of neutrino masses enjoying an improvement of 65\% over  the case with only $\fsig$ data. We find also a significant increase in the precision on the background and perturbation Horndeski parameters, with an additional gain of $\sim\! 20 \%$ for the varying effective DE equation of state parameter $w_a$ and $\sim\! 80 \%$ for the evolution of the effective Planck mass $a_M$. Extending the redshift of the mocks up to $z_{\rm max}=2$ shows that the constraints provided by the combined $\fsig$ data are already matched by the split data $f$ and $\sig$ with $z_{\rm max}=1$. 

Our results highlight that growth data, whether split or combined, with 1\% relative error could lead to the detection of deviations from $\Lambda$CDM with strong significance (more than 5$\sigma$), should the underlying cosmology be close to the current Horndeski best-fit fiducial.

The splitting of growth data on $\fsig$ into data on $f$ and $\sig$ with galaxy-galaxy lensing \cite{delaTorre:2016rxm,Shi:2017qpr,Jullo:2019lgq} or by combinations with the bispectrum \cite{Gil-Marin:2016wya} emerges clearly from this work as both a powerful complementary probe for the standard model and a stringent probe to detect departures from it. The latter could prove crucial in the era of future surveys, given the current tensions within the standard model and the emergence of alternative models of gravity favoured via Bayesian evidence \cite{Peirone:2019aua,Sola:2019jek}.

\section*{Acknowledgements}

We are grateful to Julien Bel for helpful exchanges and feedback on the draft. We thank Matteo Martinelli for useful discussions. The authors acknowledge the Sciama High Performance Compute Cluster, which is supported by the ICG at the University of Portsmouth, and the CHPC the Centre for High Performance Computing (CHPC), South Africa, for providing computational resources for this research project. LP and RM are supported by the South African Radio Astronomy Observatory (SARAO) and the National Research Foundation (Grant No. 75415). RM is also supported by the UK STFC Consolidated Grant ST/N000668/1. SI was supported by the European Structural and Investment Fund and the Czech Ministry of Education, Youth and Sports (Project CoGraDS - CZ.02.1.01/0.0/0.0/15\_003/0000437). AdlCD acknowledges financial support from NRF Grants No.120390, Reference: BSFP190416431035 and No.120396, Reference: CSRP190405427545, Project No. FPA2014-53375-C2-1-P from the Spanish Ministry of Economy and Science, Project No. FIS2016-78859-P from the European Regional Development Fund and Spanish Research Agency (AEI), and support from Projects Nos. CA15117 and CA16104 from COST Actions EU Framework Programme Horizon 2020.

\bibliography{References.bib}
\end{document}